\documentstyle[fleqn]{tp}
\input epsf
\begin{document}

\twocolumn[
\title{Polarization of the CMB Anisotropy}
\author{Martin White\\
{\it Departments of Astronomy and Physics, University of Illinois}\\
{\it at Urbana-Champaign, Urbana, IL 61801-3080, U.S.A.}}
\vspace*{16pt}   

ABSTRACT.\
I review why we expect the CMB anisotropy to be polarized,
what we can learn from studying polarization and the level
of the experimental challenge it presents.  A discussion of
current and future polarization sensitive experiments will
focus on the expected sensitivity of PLANCK.
\endabstract]

\markboth{Martin White}{Polarization of the CMB Anisotropy}

\small

\section{Introduction}

In this contribution I discuss the theoretical predictions, and
experimental prospects for detection of polarization in the anisotropy
of the cosmic microwave background (CMB) radiation.  While there is
every reason to believe that the anisotropy is polarized, there is no
experimental verification of this prediction yet.  This is not too
surprising, since the level of the polarization is but a small fraction
of the already extremely small anisotropy itself.

The degree of (linear) polarization is directly related to the quadrupole
anisotropy in the photons when they last scatter, at $z\sim 10^3$.
While the exact properties of the polarization depend on the mechanism for
producing the anisotropy, several general properties arise.
The polarization peaks at angular scales smaller than the horizon at last
scattering due to causality.
Furthermore, the polarized fraction of the temperature anisotropy is small
since only those photons that last scattered in an optically thin region
could have possessed a quadrupole anisotropy.  
The fraction depends on the duration of last scattering.
For the standard thermal history, it is $10$ per cent on a characteristic
scale of tens of arc minutes.
Since temperature anisotropies are at the $10^{-5}$ level, the polarized
signal is at (or below) the $10^{-6}$ level, or several $\mu $K, representing
a significant experimental challenge. 
However, as I shall describe below, there are many things that a study of
the polarization can teach us, so the experimental investment is well worth
while.

The outline is as follows.  In the next section I discuss why we expect
the CMB anisotropy to be (linearly) polarized.  I follow this with some
reasons why we should attempt to study this polarization, and end with
a discussion of the experimental prospects, focusing specifically on the
Planck Surveyor satellite mission.
The goal here is to provide a simple picture of the various issues involved.
For the mathematical formalism, and much more detail, the reader is referred
to (Bond \& Efstathiou~1984, Polnarev~1985, Zaldarriaga \& Seljak~1997,
Kamionkowski, Kosowsky \& Stebbins~1997, Hu \& White~1997ab).

\section{Why is the CMB supposed to be polarized?}

The Thomson scattering cross section depends on polarization as
\begin{equation}
  {d\sigma_T\over d\Omega} \propto |\hat{\epsilon}
  \cdot \hat{\epsilon}'|^2\,,
\end{equation}
where $\hat{\epsilon}$ ($\hat{\epsilon}'$) are the incident (scattered)
polarization directions.  Heuristically, the incident light sets up
oscillations of the target electron in the direction of the electric
field vector $\vec{E}$, i.e.~the polarization.
The scattered radiation intensity thus peaks in the direction normal to,
with polarization parallel to, the incident polarization.
More formally, the polarization dependence of the cross section is
dictated by electromagnetic gauge invariance and thus follows from very
basic principles of fundamental physics.

\begin{figure}
\centering\mbox{\epsfxsize=5cm \epsfbox{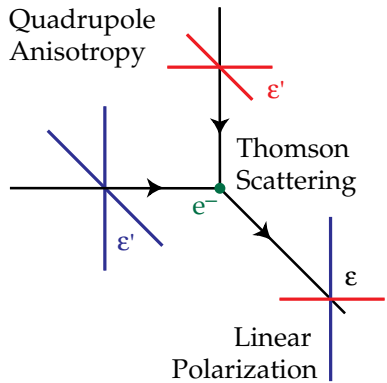}}
\caption[]{Thomson scattering of radiation with a quadrupole anisotropy
generates linear polarization.  Thick lines represent hot and thin lines
cold radiation.}
\label{fig:thomson}
\end{figure}

\begin{figure*}
\centering\mbox{\epsfxsize=9cm \epsfbox{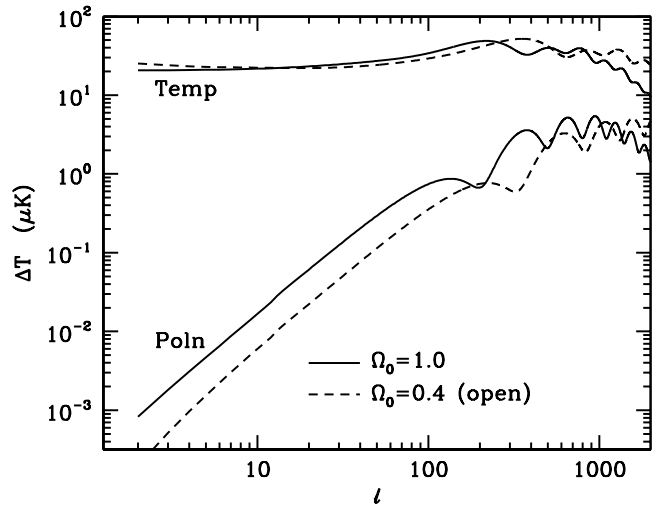}}
\caption[]{The temperature and polarization angular power spectra predicted
in two cold dark matter models, one with critical density and one open.
The model contains only density perturbations, so only the $E$-mode
polarization is non-zero.  This model assumes no late reionization of the
universe, so the large-angle polarization is purely a projection of
smaller scale polarization generated during last scattering.}
\label{fig:deltat}
\end{figure*}

If the incoming radiation field were isotropic, orthogonal polarization
states from incident directions separated by $90^\circ$ would balance so
that the outgoing radiation would remain unpolarized.   Conversely,
if the incident radiation field possesses a {\it quadrupolar\/} variation
in intensity or temperature (which possess intensity peaks at $90^\circ=\pi/2$
separations), the result is a {\it linear\/} polarization of the scattered
radiation (see Figure~\ref{fig:thomson}).
A reversal in sign of the temperature fluctuation corresponds to a
$90^\circ$ rotation of the polarization, which reflects the spin-2
nature of polarization.

The radiative transfer of polarized light in the expanding universe can be
solved numerically by following the Boltzmann equation for brightness and
polarization perturbations.  Such machinery is not necessary in order to
understand the main physical points however.
We describe polarized light by the Stokes parameters: $I$, $Q$, $U$ and
$V$.  The latter, describing circular polarization, is expected to be absent
in the cosmological context due to parity conservation.  This leaves us with
3 observables.  The intensity fluctuations are seen by us as temperature
perturbations.  It is convenient to construct a linear combination of the
$Q$ and $U$ stokes parameters.  The new basis vectors are called $E$- and
$B$-mode polarization (not to be confused with the electric and magnetic
fields of the e-m radiation itself).  The $E$-mode polarization is correlated
with the temperature, while the $B$-mode polarization is not.
Again for reasons of parity, the density perturbations which give rise to
large-scale structure in the universe generate purely $E$-mode polarization
(in the absence of gravitational lensing effects --
see Zaldarriaga \& Seljak~1998).

The temperature and $E$-mode polarization predicted in cold dark matter
models of structure formation are shown in Figure~\ref{fig:deltat}.
The vertical axis is the rms fluctuation as a function of angular scale,
where the horizontal axis is the multipole number $\ell\sim\theta^{-1}$
with $1^\circ$ corresponding to $\ell\sim 10^2$.
Note that the polarization peaks at smaller angular scales, higher $\ell$,
than the anisotropy and at about $10$ per cent of the amplitude.  A closer
examination will reveal that the peaks in the polarization spectrum are out
of phase with those in the temperature spectrum, and slightly ``sharper''.

Of critical importance is the somewhat obvious fact that polarization
is only generated by {\it scattering\/}.  Gravitational interactions do not
generate any polarization.  Thus the generation of polarization is localized
in time, at the last scattering epoch and perhaps at low-$z$ when the
universe reionized.

\section{Why should we care?}

Why should we be concerned with the polarization of CMB anisotropies?
There are 3 main reasons.  First, that the CMB anisotropies are polarized
is a fundamental prediction of the gravitational instability paradigm.
Under this paradigm, small fluctuations in the early universe grow into the
large scale structure we see today.
If the temperature anisotropies we observe are indeed the result of primordial
fluctuations, their presence at last scattering would polarize the CMB
anisotropies themselves.
The verification of the (partial) polarization of the CMB on small scales would
thus represent a fundamental check on our basic assumptions about the behavior
of fluctuations in the universe, in much the same way that the redshift
dependence of the CMB temperature is a test of our assumptions about the
background cosmology.

Furthermore, observations of polarization provide an important tool for
reconstructing the model of the fluctuations from the observed power spectrum
(as distinct from fitting an {\it a priori} model prediction to the
observations).
The polarization probes the epoch of last scattering {\it directly\/} as
opposed to the temperature fluctuations which may evolve between last
scattering and the present.  This localization in time is a very powerful
constraint for reconstructing the sources of anisotropy.
Moreover, different sources of temperature anisotropies (scalar, vector and
tensor) give different patterns in the polarization: both in its intrinsic
structure and in its correlation with the temperature fluctuations themselves.
For example, the relative prominence of the $B$- and $E$-modes of the
polarization and the slope of the spectra at low-$\ell$ distinguish the
different types of fluctuations (Hu \& White 1997ab).
A large $B/E$ ratio indicates the presence of vector modes.  Since vector
modes decay with the expansion of the universe, this tells us that whatever
forms them (e.g.~cosmological defects) must be acting now.
In an inflationary model the perturbations were laid down at early times,
and the vector modes will have decayed away by the present.
The structure of the polarization spectrum around $\ell\sim10^2$ also allows
us to unambiguously distinguish adiabatic models (e.g.~inflation) from
isocurvature models (Hu, Spergel \& White~1997, Spergel \& Zaldarriaga~1997).
Thus by including polarization information, one can distinguish the
ingredients which go to make up the temperature power spectrum and so the
cosmological model (for a detailed discussion see Hu \& White~1997b).

Finally, the polarization power spectrum provides information complementary
to the temperature power spectrum even for ordinary (scalar or density)
perturbations.
This can be of use in breaking parameter degeneracies and thus constraining
cosmological parameters more accurately.  The prime example of this is the
degeneracy, within the limitations of cosmic variance, between a change in
the normalization and an epoch of ``late'' reionization.

\begin{figure}
\centering\mbox{\epsfxsize=5cm \epsfbox{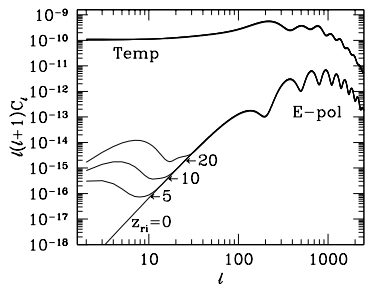}}
\caption[]{Late reionization generates large-angle polarization, allowing
the degeneracy between a change in amplitude and reionization to be broken.}
\label{fig:polz}
\end{figure}

\begin{figure}
\centering\mbox{\epsfxsize=5cm \epsfbox{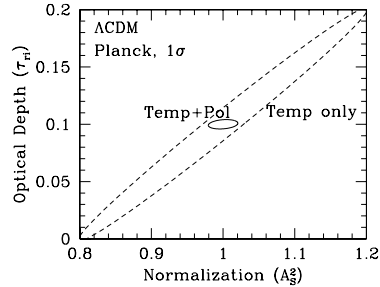}}
\caption[]{The inclusion of polarization information breaks the degeneracy
between a change in the normalization and a change in the optical depth
to Compton scattering (i.e.~the redshift of reionization).
The contours denote $1\sigma$ error ellipses in the normalization--reionization
plane, that would be determined by Planck using only temperature information
(dashed) or temperature and polarization information (solid).  All other
parameters have been marginalized over.}
\label{fig:degen}
\end{figure}

A period of late reionization re-couples the CMB photons to the matter.
The scattering that occurs erases anisotropy on scales smaller than the
horizon at that epoch, but also produces more (large-angle) polarization.
This is shown in Figure~\ref{fig:polz}.  If the reionization occurs late
enough the suppression of anisotropy is uniform for almost all multipoles,
and looks much like a change in the amplitude of the spectrum.  The
differences are constrained to large angles, small $\ell$, where cosmic
variances limits the precision to which the spectrum can be measured.
This makes these two parameters highly degenerate: an increase in the
normalization can be counteracted by an increase in the redshift of
reionization.
Because the large-angle polarization is so sensitive to reionization,
it can break this degeneracy as shown in Figure~\ref{fig:degen}.

Perhaps more important than breaking degeneracies in parameterized models,
polarization holds the key to reconstructing the underlying model for the
fluctuations directly from the observed anisotropy spectra.

\section{Experimental Prospects}

Theoretically therefore, the case for observing the polarization is very
strong.  Unfortunately the experimental challenge is daunting.
Existing upper limits are nearly an order of magnitude above the theoretical
predictions (Hu \& White~1997b).
In addition to the challenge in raw sensitivity, the foregrounds are not
well understood at any CMB frequency (Keating et al.~1998).  Free-free
emission from Thomson scattering in H{\sc ii} regions leads to polarization
of the order of 10 per cent.  Extrapolations of dust emission from higher
frequencies suggest a polarization of a few to 10 per cent.  Radio point
sources can be 20 per cent polarized.
Synchrotron radiation is perhaps the largest worry, since it can be up to
75 per cent polarized.

The answer to all of these concerns is of course the same as for the
temperature anisotropies themselves.  One observes as much of the sky as
possible, in as many wavebands as possible at the highest angular resolution
and with the highest sensitivity one can achieve.
These are the design drivers for the Planck Surveyor satellite.

\begin{figure}
\centering\mbox{\epsfxsize=5cm \epsfbox{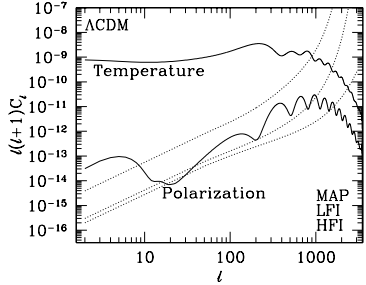}}
\caption[]{The predicted signal (solid) and noise (dashed) levels for
cosmological polarization in a $\Lambda$CDM model.  The signal level for
the temperature anisotropy is shown for reference.
The noise power spectra are for {\sl MAP\/}, the Planck LFI and HFI
respectively (top to bottom) and assume averaging into 5 per cent bands
in $\ell$.}
\label{fig:noise}
\end{figure}

\begin{figure}
\centering\mbox{\epsfxsize=5cm \epsfbox{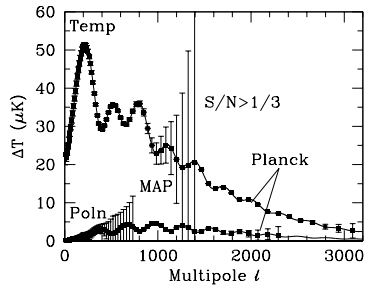}}
\caption[]{The $1\sigma$ errors on a series of uncorrelated band powers,
as would be measured with MAP and Planck.  For clarity, only points with
$S/N>1/3$ are plotted.}
\label{fig:band}
\end{figure}

In the {\it absence\/} of foregrounds the sensitivity that
{\sl MAP\/}\footnote{http://map.gsfc.nasa.gov},
the Planck\footnote{http://astro.estec.esa.nl/Planck} LFI and HFI will
achieve is shown in Figure~\ref{fig:noise}.
This figure assumes uniform coverage of the sky.  In reality Planck will
go a factor of 2 deeper in the polar caps, increasing the signal to noise
in those regions over the mean value shown in the Figure.
The signal-to-noise ratio is interpreted in a more familiar form in
Figure~\ref{fig:band}.

\section{Conclusions}

I have argued that if the structure we see did grow from initially small
perturbations, the CMB anisotropy should be polarized.  Detection of this
polarization represents a fundamental test of our theories of structure
formation.
If the CMB is polarized, then we gain two additional observables, in addition
to the temperature fluctuations themselves.  Since the temperature and $E$-mode
polarization are predicted to be correlated, this takes us from one power
spectrum ($TT$) to 4: $TT$, $TE$, $EE$ and $BB$.
The polarization power spectra have a different dependence on the cosmological
parameters than does the temperature spectrum, allowing us to break some
parameter degeneracies.

The angular spectrum of the polarization is predicted to be ``sharper''
than the temperature spectrum, to peak at smaller angular scales than the
temperature, and be $\sim 10$ per cent of its amplitude.  Measurements of
the polarization spectra provide an important cross check on our modeling
assumptions.
In principle, observations of the polarization could allow us to decompose
the model of structure formation into its building blocks, allowing full
reconstruction rather than simple model fitting.

%%%% The star below makes the section heading appear without 
%%%% a section number.  
\section*{Acknowledgments}

I would like to thank Wayne Hu for a long and productive collaboration
upon which this work is based.  Thanks also to the organizers for making
the conference the success it was.

%%%% If you have more than 99 references, but fewer than 1000, 
%%%% then you need to change 99 to 999.  

\end{document}